\documentclass[aps,prl,showpacs,twocolumn,superscriptaddress]{revtex4-1}
\usepackage{amsmath,amssymb,physics}
\usepackage{graphicx}
\usepackage{dsfont}
\usepackage{bm}
\usepackage{xcolor}

\usepackage{epstopdf}
\usepackage{hyperref}

\newcommand{\rhoL}{\rho_{\rm\scriptscriptstyle L}}
\newcommand{\rhoR}{\rho_{\rm\scriptscriptstyle R}}
\newcommand{\rhoB}{\rho_{\rm\scriptscriptstyle b}}

\DeclareMathOperator*{\argmin}{argmin}
\DeclareMathOperator*{\argmax}{argmax}

\begin{document} 
\title[]{Brownian asymmetric simple exclusion process}

\author{Dominik Lips} 
\email[]{dlips@uos.de} 
\affiliation{Universit{\"a}t Osnabr{\"u}ck, Fachbereich Physik,
  Barbarastra{\ss}e 7, D-49076 Osnabr{\"u}ck, Germany}

\author{Artem~Ryabov}
\email[]{rjabov.a@gmail.com}
\affiliation{Charles University, Faculty of Mathematics and Physics,
  Department of Macromolecular Physics, V Hole\v{s}ovi\v{c}k\'ach 2,
  CZ-18000 Praha 8, Czech Republic}

\author{Philipp Maass}
\email[]{maass@uos.de}
\affiliation{Universit{\"a}t Osnabr{\"u}ck, Fachbereich Physik,
  Barbarastra{\ss}e 7, D-49076 Osnabr{\"u}ck, Germany}

\date{June 29, 2018, revised August 31, 2018} 

\begin{abstract}
We study the driven Brownian motion of hard rods in a one-dimensional cosine
potential with an amplitude large compared to the thermal energy.  In
a closed system, we find surprising features of the steady-state
current in dependence of the particle density. The form of the
current-density relation changes greatly with the particle size and
can exhibit both a local maximum and minimum.  The changes are caused
by an interplay of a barrier reduction, blocking and exchange symmetry
effect.  The latter leads to a current equal to that of
non-interacting particles for a particle size commensurate with the
period length of the cosine potential. For an open system coupled to
particle reservoirs, we predict five different phases of
non-equilibrium steady states to occur.  Our results show that the
particle size can be of crucial importance for non-equilibrium phase
transitions in driven systems.  Possible experiments for demonstrating
our findings are pointed out.
\end{abstract}

\maketitle  

A minimal model for studying fundamental questions of statistical
physics out of equilibrium is the asymmetric simple exclusion process 
(ASEP) \cite{Derrida:1998, Schuetz:2001}, which, due to its simplicity, is sometimes referred
to as the "Ising model of non-equilibrium statistical mechanics"
\cite{Schmittmann/Zia:1995}.  In this model, particles with exclusion
interaction are considered to hop between neighboring sites of a
one-dimensional lattice with a bias in one direction. Many intriguing
findings were reported for this ASEP, as exact results for microstate
distributions in non-equilibrium steady states (NESS)
\cite{Blythe/Evans:2007}, phase transitions of NESS in
systems with open boundaries \cite{Krug:1991, Parmeggiani/etal:2003}
and defects \cite{Kolomeisky:1998}, condensation transitions in
systems with random \cite{Evans:1996} and non-Poissonian hopping rates
\cite{Concannon/Blythe:2014}, and singular points in rate functions
characterizing large deviations of fluctuations in time-averaged
currents \cite{Bertini/etal:2005, Lazarescu:2015, Baek/etal:2017}. 

Most applications of the ASEP are found in the modeling of 
biological traffic \cite{Schadschneider/etal:2010, Chou/etal:2011}, where the model was first introduced to
describe protein synthesis by ribosomes \cite{MacDonald/etal:1968} and where it is frequently used now
in studies of molecular motor motion \cite{Kolomeisky:2013, Appert-Rolland/etal:2015}. Clearly,
refinements of the core model are needed for corresponding applications, 
such as the consideration of inhomogeneous hopping rates, particles occupying several sites, internal states of
particles and multi-lane variants \cite{Schadschneider/etal:2010, Chou/etal:2011, Kolomeisky:2013, Appert-Rolland/etal:2015}. 
A direct comparison of models and experiments in this area, 
however, is difficult to realize and hampered by the complexity of biological transport phenomena. 

Here, we consider a Brownian
motion of particles with the following ingredients resembling features of the ASEP:
(i) an exclusion interaction between particles over a range $\sigma$, (ii)
a periodic potential $U(x)$ with period length $\lambda$, giving rise
to an effective hopping motion of the particles between the potential
wells, and (iii) a constant drag force $f$ acting on the particles.
This BASEP is a broadly applicable model of single-file diffusion
\cite{Hahn/etal:1996, Taloni/etal:2017}, 
and can be realized in lab by  driving colloids using a combination of microfluidics and optical micro-manipulation 
techniques \cite{Arzola/etal:2017, Skaug/etal:2018, Schwemmer/etal:2018}. 
These recent experiments have a potential to probe and 
verify fundamental theoretical predictions for non-equilibrium collective phenomena.
The lattice ASEP may, due to its discreteness, not be a correct 
model for corresponding experimentnal tests. 

\begin{figure}[b!]
\includegraphics[width=0.85\columnwidth]{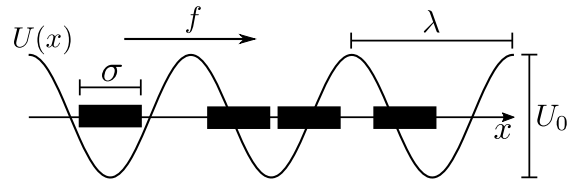}
\caption{Driven Brownian motion of interacting particles of size
  $\sigma$ in a cosine potential with barrier height $U_0$ and period
  length $\lambda$ under a drag force $f$.}
\label{fig:model} 
\end{figure}

\begin{figure*}[t!]
\includegraphics[width=1.0\textwidth]{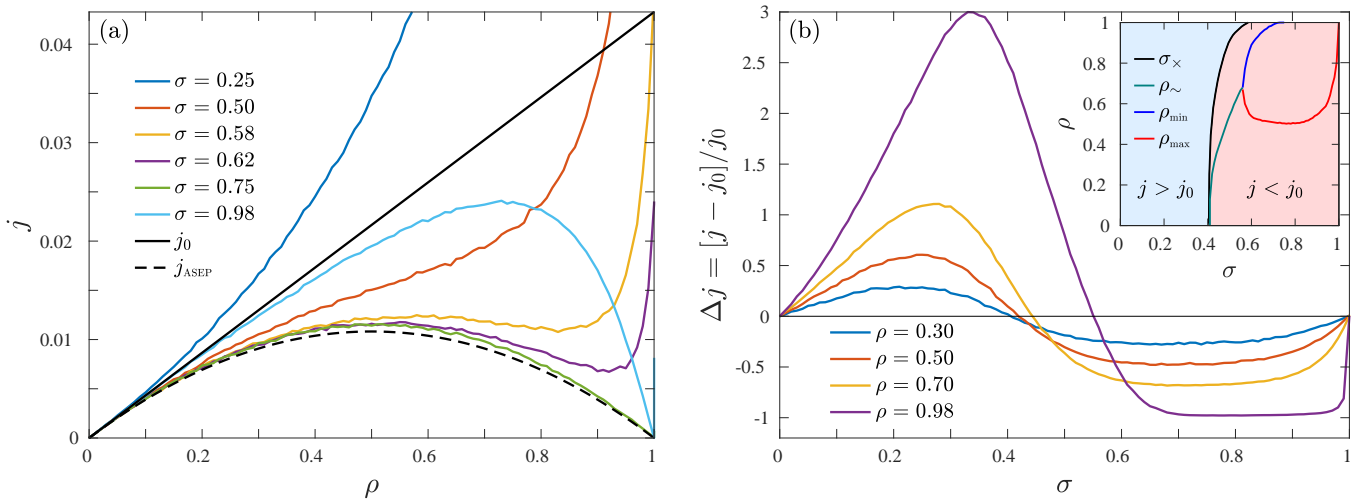}
\caption{(Color online) (a) Current-density relations for various
  fixed particle sizes $\sigma$.  The solid and dashed black lines
  mark the currents $j_0(\rho)$ and $j_{\rm\scriptscriptstyle
    ASEP}(\rho)$ for non-interacting particles and the corresponding
  ASEP, respectively.  (b) Particle size dependence of the current
  change $\Delta j(\rho,\sigma)=[j(\rho,\sigma)-j_0(\rho)]/j_0(\rho)$
  due to hard-core interactions for different fixed densities $\rho$.
  The inset shows the curve $\sigma_\times(\rho)$, which separates the
  region of current enhancement (blue area) and reduction (red area), and the dependence of
  $\rho_\sim$, $\rho_{\rm\scriptscriptstyle max}$ and
  $\rho_{\rm\scriptscriptstyle min}$ on $\sigma$.}
\label{fig:current} 
\end{figure*}

Indeed, we show in this Letter that the BASEP exhibits surprising features
which have no counterpart in the ASEP.  These features are a
consequence of the length scale $\sigma$, which enters the problem as
a parameter in addition to the particle density $\rho$.  The site
blocking effect associated with the exclusion interaction is
dominating the steady-state particle current $j(\rho,\sigma)$ in a
limited $\sigma$ range only.  Due to a barrier reduction effect,
$j(\rho,\sigma)$ can be larger than the current $j_0(\rho)$ of
non-interacting particles.  An exchange symmetry effect emerges when
$\sigma$ becomes commensurate with the period length $\lambda$.  In
this case $j(\rho,\sigma\!=\!\lambda)=j_0(\rho)$, as if there were no
interactions.  The interplay of the barrier reduction, blocking and
exchange symmetry effects leads to changes of the form of the
current-density relation with the particle size.  This in turn leads
to the appearance of five different non-equilibrium phases in open BASEPs
coupled to particle reservoirs.

Figure~\ref{fig:model} illustrates interacting particles of size
$\sigma$ that are driven by a drag force $f$ through a cosine
potential $U(x) =(U_0/2)\cos(2\pi x/\lambda)$ with barrier height
$U_0$.  Their center of mass positions $x_i$, $i=1,\ldots,N$, are
considered to perform an overdamped Brownian motion according to the
coupled Langevin equations
\begin{equation}
\frac{\dd x_i}{\dd t}=\mu\left[f^{\rm\scriptscriptstyle ext}(x_i)
+f_i^{\rm\scriptscriptstyle int}\right]+\sqrt{2D}\,\eta_i(t)\,,
\label{eq:langevin}
\end{equation}
where $f^{\rm\scriptscriptstyle ext}(x)=f-\dd U(x)/\dd x$ is the
external force, $f_i^{\rm\scriptscriptstyle
  int}=f_i^{\rm\scriptscriptstyle int}(x_1,\ldots x_N)$ is the
interaction force on the $i$th particle, and $\eta_i(t)$ are
independent Gaussian white noise processes with zero mean and
$\langle\eta_i(t)\eta_j(t')\rangle=\delta_{ij}\delta(t-t')$; $\mu$ and
$D=k_{\rm\scriptscriptstyle B}T\mu$ are the bare mobility and
diffusion coefficient, respectively, and $k_{\rm\scriptscriptstyle
  B}T$ is the thermal energy. In the BASEP,
$f_i^{\rm\scriptscriptstyle int}$ is solely determined by the
hard-core exclusion between neighboring particles, i.e.\ a
contribution upon particle contact. The system size $L$ is taken to be
an integer multiple of $\lambda$ and periodic boundary conditions are
imposed. As units for length, time, and energy we choose $\lambda$,
$\lambda^2/D$, and $k_{\rm\scriptscriptstyle B}T$, respectively. The
density, or filling factor, is $\rho=N/L$.  We set $U_0\gg
k_{\rm\scriptscriptstyle B}T$ to generate an effective hopping motion
of the particles, and focus first on the case where both $\rho$ and $\sigma$
lie in the range $[0,1]$. 

To determine $j(\rho,\sigma)$ in the non-equilibrium steady state
(NESS), we have carried out Brownian dynamics simulations, which we
corroborate by analytical considerations.  The barrier height and the
drag force are fixed by setting $U_0=U_0/(k_{\rm\scriptscriptstyle
  B}T)=6$ and $f=f\lambda/(k_{\rm\scriptscriptstyle B}T)=1$.  In most
of the simulations we have chosen $L=100$. For $\rho$ and $\sigma$
close to one, simulations were performed also for larger $L$ to check
that our results are not affected by the finite system size.  The
hard-core interaction force between neighboring particles was
simulated according to the algorithm developed in
\cite{Behringer/Eichhorn:2012}. For $\sigma$ close to one, we also
used the method proposed in \cite{Scala/etal:2007}.

For non-interacting particles, the current increases linearly with
$\rho$, $j_0=v_0\rho$, where $v_0$ is the mean velocity of a single
particle and can be calculated analytically
\cite{Ambegoakar/Halperin:1969}; for our parameters $v_0\cong0.043$. By
the hard-core interaction, this linear current-density relation is
modified in quite different ways for different particle sizes
$\sigma$, as can be seen from Fig.~\ref{fig:current}(a).  As reference
curves, we included in this figure the line $j_0=v_0\rho$ for
non-interacting particles (solid black line) and the corresponding one
for the ASEP \cite{Derrida:1998, Schuetz:2001},
$j_{\rm\scriptscriptstyle ASEP}(\rho)=j_0(\rho)(1-\rho)=v_0\rho(1-\rho)$
(dashed line). Remarkably, the parabolic curve of the ASEP is resembled
in a quite limited $\sigma$ range only.

To understand the nonlinear current-density relation for different
particle sizes, it is helpful to first consider the relative current
change $\Delta j(\rho,\sigma)=[j(\rho,\sigma)-j_0(\rho)]/j_0(\rho)$
due to the interactions as a function of $\sigma$ for several fixed
$\rho$. Corresponding curves plotted in Fig.~\ref{fig:current}(b) show
a similar behavior for all $\rho$. For small $\sigma$, $\Delta j$
increases with $\sigma$ up to a maximum and then it decreases until
crossing the zero line at a value $\sigma_\times(\rho)$. Hence, for
$0<\sigma<\sigma_\times(\rho)$, $j(\rho,\sigma)$ becomes enhanced
compared to $j_0(\rho)$.  When increasing $\sigma$ beyond
$\sigma_\times(\rho)$, $\Delta j$ first decreases, then remains
approximately constant in a plateau-like regime, and eventually
increases again, where $\Delta j=0$ for $\sigma=1$ and all particle
densities $\rho$.  Hence, $j(\rho,\sigma)$ becomes reduced compared to
$j_0(\rho)$ for $\sigma_\times(\rho)<\sigma<1$, and it becomes equal
to $j_0$ for $\sigma=1$.  The crossover value $\sigma_\times(\rho)$
increases with $\rho$ and the full curve shown in the inset of
Fig.~\ref{fig:current}(b) divides the $\sigma$-$\rho$-plane in two
regions of current enhancement and reduction.

\begin{figure}[t!]
\includegraphics[width=\columnwidth]{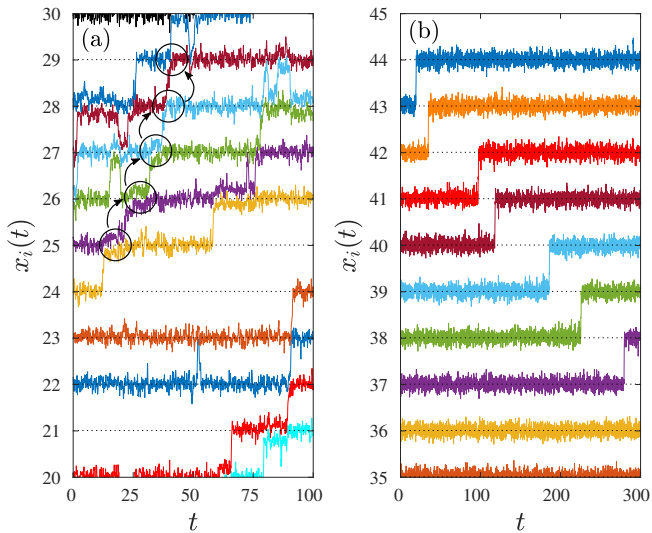}
\caption{(Color online) Typical particle trajectories in the BASEP for $\rho=0.75$,
   and (a) $\sigma=0.25$ and (b) $\sigma=0.75$.
  The horizontal dotted lines indicate the positions of
  the potential minima.  In (a) a cascade-like propagation of
  double-occupancies marked by the circles is demonstrated by the
  arrows. In (b) potential wells are vacated and filled in a
  sequential process.}
\label{fig:trajectories}
\end{figure}

The enhancement of the current is caused by a barrier reduction effect, which occurs if
a potential well is occupied by more than one particle \footnote{A well is $n$ times occupied, if exactly
$n$ of the particles' center positions lie in the $x$-interval between the successive 
potential maxima enclosing the well.}. Inside a multi-occupied well, the mutually excluding
particles exhibit, on average, higher potential energies than a particle in a single-occupied well.
They need to surmount a lower barrier for escaping the well, which causes the current enhancement. 
This enhancement is the stronger the larger $\rho$ [see Fig.~\ref{fig:current}(b)],
because the probability of multi-occupancies rises with increasing $\rho$. Also, 
clusters of neighboring occupied wells become larger on average. This facilitates a cascade-like
propagation of multi-occupations as demonstrated in Fig.~\ref{fig:trajectories}(a).

The enhancement effect due to barrier reduction is pronounced at small
$\sigma$, because for larger $\sigma$, the formation of double (or
higher) occupancies requires larger energies and becomes less likely.
For $\sigma>\sigma_\times(\rho)$, the blocking effect, known from the
ASEP, prevails.  It means that an effective hopping of a particle to a
neighboring well is suppressed if the target well is occupied. Typical
particle trajectories in this regime, as displayed in
Fig.~\ref{fig:trajectories}(b), show that clusters of neighboring 
particles frequently move in a manner, where wells are sequentially
vacated and filled. Hence, when increasing $\sigma$ beyond
$\sigma_\times(\rho)$, the particle motion becomes similar to a
hopping on a lattice with forbidden multi-occupation of sites. The
current then is nearly independent of the particle size $\sigma$, as
reflected in the plateau-like $\sigma$-intervals in
Fig.~\ref{fig:current}(b).

To understand why the current increases again for large $\sigma$
approaching one, let us consider a coordinate transformation $x_i\to
x_i'=x_i-i\sigma$ in the Langevin equations (\ref{eq:langevin}),
which for $\sigma=\lambda=1$ leaves them invariant, because of the
$\lambda$-periodicity of $f^{\rm\scriptscriptstyle ext}(x)$. After
this transformation, the dynamics of the $x_i'$ corresponds to
that of point particles. However, for point particles with hard-core
interaction, collective properties, like the current, become invariant
under particle exchange \cite{Ryabov/Chvosta:2011} and accordingly
$j(\rho,1)=j_0(\rho)$ for all $\rho$. 

Refining this line of reasoning, we show in the supplemental material \cite{suppl} that
the current for general $\sigma\ge0$ with 
$m=\mathrm{int}(\sigma/\lambda)$ fulfils the relation
\begin{equation}
j(\rho,\sigma)=(1-m\rho)\,j\left(\frac{\rho}{1-m\rho},\sigma-m\lambda\right)\,.
\label{eq:j-relation}
\end{equation}
This relation means that the current behavior for $\sigma\ge\lambda$ can be inferred from
that for $\sigma<\lambda$. Moreover, it implies:
(i) $j(\rho,\sigma)=j_0(\rho)$ for all $\sigma=m\lambda$,
$m=1,2,\ldots$; (ii) $j(\rho,\sigma)$ can resemble the behavior of an $l$-ASEP \cite{Lakatos/Chou:2003}, 
where particles occupy $l$ lattice sites.

All results are further supported by analytical calculations when
starting from the Smulochowski equation for the joint probability
density $p_N(x_1,\ldots,x_N,t)$ of finding the particles at positions
$x_1,\ldots,x_N$ at time $t$ \cite{suppl}.  
In the NESS, the current $j(\rho,\sigma)$ is given by
\begin{align}
j(\rho,\sigma)&=\left[\mu\bigl(f^{\rm\scriptscriptstyle ext}(x)+\langle f^{\rm\scriptscriptstyle int}(x)\rangle\bigr) 
-D\frac{\dd}{\dd x}\right] \rho_{\rm\scriptscriptstyle loc}(x)\,,
\label{eq:j1}
\end{align}
where $\rho_{\rm\scriptscriptstyle loc}(x)$ is the local density and
\begin{align}
\langle f^{\rm\scriptscriptstyle int}(x)\rangle&
=k_{\rm\scriptscriptstyle B}T\left[\psi_-(x)-\psi_+(x)\right]
\label{eq:barf}
\end{align}
is the mean interaction force on a particle at position $x$. Here,
$\psi_-(x)=\Psi_-(x)/\rho_{\rm\scriptscriptstyle loc}(x)$ and
$\psi_+(x)=\Psi_+(x)/\rho_{\rm\scriptscriptstyle loc}(x)$ are the
conditional probability densities that, given a particle at position
$x$, a neighboring particle is in contact (at distance $\sigma$) to it
in counterclockwise and clockwise direction, respectively;
$\Psi_\pm(x)$ are the respective joint probability densities in the
NESS. Because $\Psi_-(x)=\Psi_+(x)$ for $\sigma=\lambda$, one obtains
$\langle f^{\rm\scriptscriptstyle int}(x)\rangle=0$ for
$\sigma=\lambda=1$ from Eq.~(\ref{eq:barf}), and it follows
$j(\rho,1)=j_0(\rho)$ from Eq.~(\ref{eq:j1}).

Moreover, multiplying Eq.~(\ref{eq:j1}) with
$1/\rho_{\rm\scriptscriptstyle loc}(x)$ and integrating over one
period $\lambda$, we obtain, when utilizing the $\lambda$-periodicity
of $\rho_{\rm\scriptscriptstyle loc}(x)$ in the NESS and of $U(x)$ in
$f^{\rm\scriptscriptstyle ext}(x)=f-\dd U(x)/\dd x$,
\begin{equation}
j(\rho,\sigma)
=\frac{\mu(f+\bar f^{\rm\scriptscriptstyle int})\lambda}{\displaystyle \int_0^\lambda\,
\frac{\displaystyle \dd x}{\displaystyle \rho_{\rm\scriptscriptstyle loc}(x)}}\,.
\label{eq:j}
\end{equation}
where $\bar f^{\rm\scriptscriptstyle int}=
\lambda^{-1}\int_0^\lambda\dd x \langle f^{\rm\scriptscriptstyle
  int}(x)\rangle$ is the period-averaged mean interaction force.  The
form of this exact expression for the current is fully analogous to
the corresponding one for a single particle
\cite{Ambegoakar/Halperin:1969}, but here it refers to a many-body
system with hard-core interactions, where $\bar
f^{\rm\scriptscriptstyle int}$ gives an additional contribution to the
driving force and is related to the two-particle density in the NESS
via Eq.~(\ref{eq:barf}). In fact, Eq.~(\ref{eq:j}) is valid also for
other interaction forces if the corresponding mean interaction force
is used. We further demonstrate \cite{suppl} that an
approximate analysis of Eq.~(\ref{eq:j}) in the linear response limit
reproduces qualitatively the behavior shown in
Figs.~\ref{fig:current}(a) and (b).

\begin{figure}[t!]
\includegraphics[width=\columnwidth]{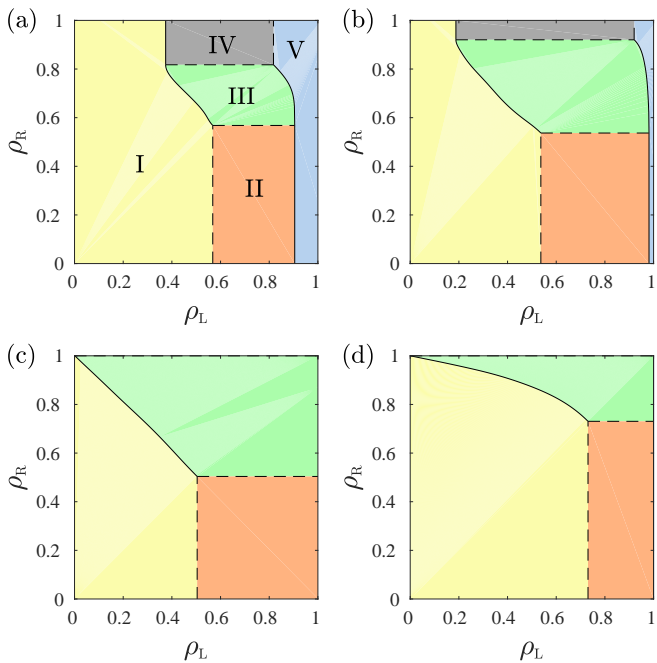}
\caption{(Color online) Phase diagrams of the open BASEP for particle sizes
(a) $\sigma=0.58$, (b) $\sigma=0.62$, (c) $\sigma=0.75$, and (d) $\sigma=0.98$.
Dashed and solid lines indicate phase transitions of first and second order, respectively.
The regions labelled I-V in (a) mark two left-boundary induced phases (I and V), a right-boundary induced
phase (III), a maximal current phase (II), and a minimal current phase (IV).
These phases are equally colored in all graphs.} 
\label{fig:phase-diagrams} 
\end{figure}

Let us now discuss the curves in Fig.~\ref{fig:current}(a) with increasing $\sigma$.
For small $\sigma=0.25$, $j(\rho,\sigma)$ is larger than $j_0(\rho)$ and increases 
monotonically due to the barrier reduction effect.
Enlarging $\sigma$, the blocking effect
becomes more relevant, which causes the curves to approach
more and more $j_{\rm\scriptscriptstyle ASEP}(\rho)$.  First, this leads to a change of curvature
of $j(\rho,\sigma)$ from concave to convex at a density $\rho=\rho_\sim$, see the curve for
$\sigma=0.5$. Then, when $\sigma$ exceeds
a critical value $\sigma_c\cong0.55$, 
a local maximum at $\rho=\rho_{\rm\scriptscriptstyle max}$ and a 
local minimum at  $\rho=\rho_{\rm\scriptscriptstyle min}$ occurs, see the curves for
$\sigma=0.58$ and $\sigma=0.62$. Upon further increasing $\sigma$,
a range of particle sizes
appears, where $j(\rho,\sigma)\simeq j_{\rm\scriptscriptstyle ASEP}(\rho)$.
Eventually the exchange symmetry effect becomes noticeable,
which causes $j(\rho,\sigma)$ to approach $j_0(\rho)$. Going along with this is a
shift of the local maximum $\rho_{\rm\scriptscriptstyle max}$ and of
$j(\rho_{\rm\scriptscriptstyle max},\sigma)$ towards larger values, see the curve for $\sigma=0.98$.
The dependence of $\rho_\sim$, $\rho_{\rm\scriptscriptstyle max}$ and 
$\rho_{\rm\scriptscriptstyle min}$ on $\sigma$ is shown in the inset of Fig.~\ref{fig:current}(b).

The different forms of the current-density relation lead to a
versatile emergence of NESS phases in an open BASEP in contact with
two particle reservoirs at its left and right end.  In this open
BASEP, the period averaged densities
$\bar\rho_i=\lambda^{-1}\int_{(i-1)\lambda}^{i\lambda}\dd
x\rho_{\rm\scriptscriptstyle loc}(x)$ in each well
$i=1,\ldots,L/\lambda$ are no longer equal in the NESS, but approach a
constant ``bulk value'' $\rho_{\rm b}$ in the system's interior far
from the boundaries.  This bulk density $\rho_{\rm b}$ or its
derivative can change abruptly upon variation of the system-reservoir
couplings, i.e.\ the parameters controlling the particle exchange with
the reservoirs.  The corresponding sets of phase transition points
separate NESS phases, in which the order parameter $\rho_{\rm b}$
varies smoothly with the system-reservoir couplings.

Independent of the details of the couplings, all possible NESS phases
can be uncovered from $j(\rho,\sigma)$ by considering just two control
parameters $\rhoL ,\rhoR\in[0,1]$, which for {\it bulk-adapted}
couplings represent the particle densities in the left and right
reservoir, respectively \cite{Hager/etal:2001, Dierl/etal:2012,
  Dierl/etal:2013}. The different phases are obtained by applying the
extremal current principles \cite{Krug:1991, Popkov/Schuetz:1999},
which state that $\rho_{\rm b}$ assumes the value at which
$j(\rho,\sigma)$ becomes minimal (for $\rhoL<\rhoR$) or maximal (for
$\rhoR<\rhoL$) in the $\rho$-intervals enclosed by $\rhoL$ and $\rhoR$:
\begin{equation}
\rho_{\rm b}=\left\{\begin{array}{l@{\hspace{1em}}l}
\displaystyle\argmin_{\rhoL\le\rho\le\rhoR}\{j(\rho,\sigma)\}\,, & \rhoL\le\rhoR\,,\\[3ex]
\displaystyle\argmax_{\rhoR\le\rho\le\rhoL}\{j(\rho,\sigma)\}\,, & \rhoR\le\rhoL\,.
\end{array}\right.
\label{eq:min-max-cur}
\end{equation}

For $\sigma<\sigma_c$, the extremal current principles in
Eq.~(\ref{eq:min-max-cur}) imply that no phase transitions occur in
the open BASEP, because $j(\rho,\sigma)$ exhibits no local minima or
maxima. For $\sigma>\sigma_c$, by contrast, phase transitions occur and
we show in Figure~\ref{fig:phase-diagrams} different examples of phase
diagrams of NESS.  Dashed and solid lines in these diagrams indicate
phase transitions of first and second order, respectively.

In Fig.~\ref{fig:phase-diagrams}(a), $\sigma=0.58$ is close to
$\sigma_c$ and in total five NESS phases appear. For the left-boundary
induced phases I and V, $\rhoB=\rhoL$, and for the right-boundary
induced phase III, $\rhoB=\rhoR$; phase II is a maximal current phase
with $\rhoB=\rho_{\rm\scriptscriptstyle max}\cong0.60$ and phase IV is a minimal
current phase with $\rhoB=\rho_{\rm\scriptscriptstyle min}\cong0.83$ [see
Fig.~\ref{fig:current}(a)].  
With increasing $\sigma$, the phase
regions II-IV extend, while the regions I and V shrink, see
Fig.~\ref{fig:phase-diagrams}(b). 
Let us note that the topology of the
phase diagrams in Figs.~\ref{fig:phase-diagrams}(a) and (b)
resembles features seen in corresponding phase diagrams
of driven lattice gases with repulsive nearest-neighbor interactions 
\cite{Popkov/Schuetz:1999, Hager/etal:2001, Dierl/etal:2013}.
For the phase diagram in Fig.~\ref{fig:phase-diagrams}(b), we
demonstrate the occurrence of different NESS phases in
simulations of an open BASEP in the supplemental material \cite{suppl}.

When entering the $\sigma$ regime,
where $j(\rho,\sigma)\simeq j_{\rm\scriptscriptstyle ASEP}(\rho)$, the
phase diagram resembles that of the ASEP with the phases I-III, as
shown in Fig.~\ref{fig:phase-diagrams}(c)\footnote{Five NESS phases as
  in Figs.~\ref{fig:phase-diagrams}(a) and (b) would appear when
  allowing for filling factors $\rho$ larger than one as a result of
  an extension of the current-density relation in
  Fig.~\ref{fig:current}(a) to the regime $1<\rho<1/\sigma$. This
  regime will be discussed elsewhere.}.  Finally, when the exchange
symmetry effect causes $\rho_{\rm\scriptscriptstyle max}$ to approach one, the phase
regions II and III shrink at the expense of region I, see
Fig.~\ref{fig:phase-diagrams}(d).

To conclude, the interplay of the barrier reduction, blocking, and
exchange symmetry effects gives rise to a surprisingly versatile form
of the current-density relation in the BASEP in dependence of the
particle size. This leads us to predict the appearance of in total
five different phases of NESS states in the open BASEP coupled to
particle reservoirs. Only for a quite limited range of particle sizes
is the behavior of the BASEP resembled by the ASEP. 

Similar non-equilibrium phase transitions as in the BASEP are expected
to occur for particles with soft repulsive interactions of short
range.  Indeed, we could identify these in simulations with a Yukawa
interaction. For a power-law soft core, just an ASEP-like behavior was
reported earlier \cite{Rodrigues/Dickman:2010}.

Besides its relevance in biology, where confined transport processes
through channels with binding sites \cite{Bauer/Nadler:2006,
  Kahms/etal:2009, Bressloff/Newby:2013} are mediated by driven
Brownian motion, we believe that the BASEP is an ideal in-situ tunable
model system for an experimental exploration of non-equilibrium phase
transitions.  Current experimental micro-manipulation techniques allow
precise engineering and fine tuning of relevant aspects of the model:
the external tilted periodic potential and the confinement. The
precise control of the potential may be achieved with high precision
using holographic optical tweezers as it was done in a related
experimental work \cite{Arzola/etal:2017}.  The confinement can be
realized within microfluidic chips. Combining microfluidics with
optical tweezers already proved to be a realizable method to probe
fundamentals of facilitated diffusion in confined spaces
\cite{Pagliara/etal:2013, Pagliara/etal:2014} and of escape times in
single-file transport \cite{Locatelli/etal:2016}. Another intriguing
recent option is to consider a nanofluidic
ratchet~\cite{Skaug/etal:2018, Schwemmer/etal:2018}, where the periodic potential landscape
is shaped by the geometry of a nanofuidic slit and an additional
electrostatic interaction between particles and walls.

\begin{acknowledgments}
This work has been funded by the Deutsche Forschungsgemeinschaft (MA
1636/10-1), the Czech Science Foundation (project no.\ 17-06716S), the
DAAD (57336032) and the M\u{S}MT (7AMB17DE014).  We sincerely thank
W.~Dieterich and the members of the DFG Research Unit FOR~2692 for
fruitful discussions.
\end{acknowledgments}


%

\end{document}